\documentstyle[aps,prl,multicol,floats,epsf]{revtex}
\begin{document}

\draft

\preprint{}

\wideabs{

\title{The irreversibility line of overdoped Bi$_{2+x}$Sr$_{2-(x+y)}$Cu$_{1+y}$O$%
_{6\pm \delta }$ at ultra-low temperatures and high magnetic fields}
\author{A. Morello$^{1,2}$, A.G.M. Jansen$^1$, R.S. Gonnelli$^2$, and S.I. Vedeneev$%
^3$}
\address{$^1$ High Magnetic Field Laboratory, Max Planck Institut f\"ur\\
Festk\"orperforschung and Centre National de la Recherche Scientifique,\\
BP166, 38042 Grenoble Cedex 9, France}
\address{$^2$ INFM - Dipartimento di Fisica, Politecnico di Torino, c.so Duca degli\\
Abruzzi 24, 10129 Torino, Italy}
\address{$^3$ P.N. Lebedev Physical Institute, Russian Academy of Sciences, SU-117924%
\\
Moscow, Russia}
\date{\today }
\maketitle

\begin{abstract}
The irreversible magnetization of the layered high-$T_{{\rm c}}$
superconductor Bi$_{2+x}$Sr$_{2-(x+y)}$Cu$_{1+y}$O$_{6\pm \delta }$
(Bi-2201) has been measured by means of a capacitive torquemeter up to $B_{%
{\rm a}}=28$ T and down to $T=60$ mK. No magnetization jumps, peak
effects or crossovers between different pinning mechanisms appear
to be present. The deduced irreversibility field $B_{{\rm irr}}$
can not be described by the law $B_{{\rm irr}}(T)\propto
(1-T/T_{{\rm c}})^n$ based on flux creep, but an excellent
agreement is found with the analytical form of the melting line of
the flux lattice as calculated from the Lindemann criterion.


\end{abstract}

}

\narrowtext

The investigation of the magnetic properties of high-$T_{{\rm c}}$
superconductors (HTS) has attracted a lot of interest, both experimental and
theoretical, thanks to the spectacular variety of new and surprising
phenomena. The research both on the flux-line statics and dynamics, which
has important technological consequences connected to the magnetic
irreversibility, and on the fundamental magnetic properties of the
superconducting state, like the upper critical field $B_{{\rm c2}}$ and the
subtle nature of the superconducting transition in magnetic field, has been
widely developed. These topics are closely related in view of the very
intriguing phenomenon of the upward curvature of the upper critical field $%
B_{{\rm c2}}(T)$ at low temperatures observed in many HTS \cite
{osof93,mack93,walk95,klei96,alex96}, which is in full contrast with the
traditional Werthamer-Helfand-Hohenberg theory \cite{wert66}. The first
suggestion that dissipative measurings tend to yield the irreversibility
line rather than the upper critical field dates more than ten years ago \cite
{malo88}. Subsequently, other models have been proposed that focus on the
influence of thermal fluctuations \cite{coop95}, vortex lattice melting \cite
{kotl96}, and low vortex viscosity \cite{gesh98} on the resistive critical
field. Other authors pointed out more radical reasons for the upward
curvature of $B_{{\rm c2}}(T)$, like the Bose-Einstein condensation of
charged bosons formed above $T_{{\rm c}}$ \cite{alex96}, the influence of
magnetic impurities \cite{ovch95}, or the role of saddle-point singularities
in the electron spectrum \cite{abri97}.

The layered cuprate superconductor Bi$_2$Sr$_2$CuO$_6$ (Bi-2201)
has been one of the first compounds showing an anomalous
temperature dependence of the resistive critical field in a thin
film \cite{osof93}. Subsequent investigations on this compound
suffered from difficulties in growing high-quality single
crystals. It is only very recently that a study of the DC
magnetization has been reported \cite{wen99}. On the other hand,
while sharing most of the structural and physical properties with
the other HTS, the low critical temperature of this superconductor
allows an experimental investigation of the whole $B-T$ phase
diagram at the currently available high magnetic fields and low
temperatures. To our knowledge, no extensive studies of the vortex
assembly in Bi-2201 have been reported so far. The study of the
vortex state in the complete phase diagram is not only interesting
in itself, but might have important implications for the anomalous
features of the upper critical field.

In this Letter, we investigate the irreversible magnetization of a
high-quality overdoped Bi$_{2+x}$Sr$_{2-(x+y)}$Cu$_{1+y}$O$_{6\pm \delta }$
single crystal. The sample was grown from a solution-melt in KCl \cite
{gori94}, and has the approximate size of $1100\times 700\times 10$ ${\rm %
\mu }$m$^3$ (mass $\sim 0.2$ mg). Both the growth method and the
small size are advantageous for an extreme homogeneity of the
sample. The intrinsic overdoping is due to the Bi excess localized
on the Sr positions. The magnetization loops were obtained by
means of torque magnetometry, using a
very sensitive capacitive torquemeter. We reached temperatures down to $%
T=60 $ mK in continuous magnetic fields $B_{{\rm a}}$ up to 28 T.
The field sweep rate d$B_{{\rm a}}/$d$t=15$ mT/s was chosen in
order to have a maximum measuring time for each loop of about one
hour with a typical thermal drift smaller than 5 - 10 mK in the
whole temperature range. Preliminary
measurements were performed in a 10 T superconducting magnet, using d$B_{%
{\rm a}}/$d$t=10.8$ mT/s. From our magnetization experiments, we evaluated a
critical temperature $T_{{\rm c}}\approx 4$ K, in agreement with the
overdoping of the sample.

Although we are interested in the irreversibility line for fields applied
perpendicular to the $ab$-planes of the crystal, the torque method has no sensitivity for $%
{\bf B}_{{\rm a}}\parallel c$. The relationship ${\bf \tau }={\bf M\times B}%
_{{\rm a}}$ between torque density ${\bf \tau }$ , magnetization ${\bf M}$
and applied magnetic field ${\bf B}_{{\rm a}}$ suggests that the torque
signal can be increased by choosing large values of the angle $\theta $
between the applied field and the $c$-axis of the sample. Actually, in the
case of a strongly two-dimensional superconductor like Bi-2201, the scaling
analysis in the large anisotropy limit of the Ginzburg-Landau model \cite
{blat92,blat94} allows to say that the magnetization ${\bf M}$ lies very
close to the $c$-axis, while its magnitude is fully determined by the
effective field $B_{{\rm a}}\cos \theta $ perpendicular to the $ab$-planes.
We have chosen $\theta =30^{\circ }$, so that the actual irreversibility
field $B_{{\rm irr}}$ for ${\bf B}_{{\rm a}}\parallel c$ is given by $B_{%
{\rm irr}}=B_{{\rm irr}}^{({\rm a})}\cos 30^{\circ }$, where $B_{{\rm irr}%
}^{({\rm a})}$ is the applied field corresponding to the vanishing of the
magnetic irreversibility. In other words, $B_{{\rm irr}}$ is the
irreversibility field that we expect to obtain in an ideal experiment with $%
B_{{\rm a}}\parallel c$ (i.e. with $\theta =0^{\circ }$), which is not
directly measurable with the torque magnetometry technique. From the torque
loops, the magnetization can be calculated as $M=\tau /(B_{{\rm a}}\sin
30^{\circ })$ but, in view of the arbitrary scaling of the measured torque
density, there is no need to take into account the $\sin 30^{\circ }$ factor.

\begin{figure}[t]
\epsfxsize=8.5cm \centerline{\epsffile{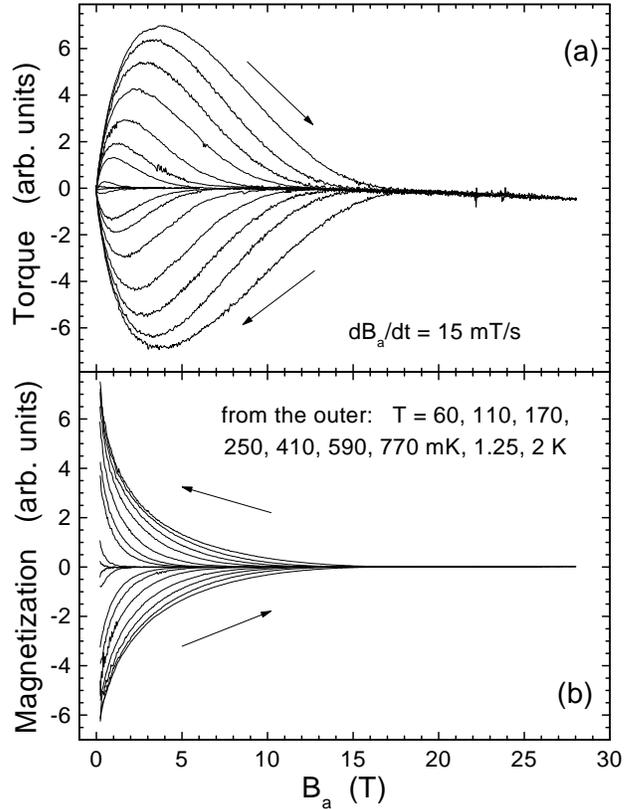}} \vspace{1mm}
\caption{The torque loops (a) and the corresponding magnetization
loops (b)
of the Bi-2201 single crystal sweeping the field up and down at d$B_{{\rm a}}/$d$%
t=15$ mT/s. Notice the absence of jumps or secondary peaks. The
magnetization is plotted only for $B_{{\rm a}}>0.2$ T.}
\label{fig:1}
\end{figure}

\begin{figure}[t]
\epsfxsize=8.5cm \centerline{\epsffile{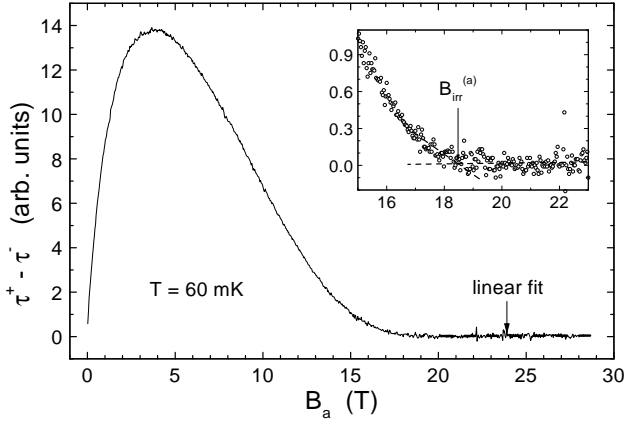}} \vspace{2mm}
\caption{The difference $\tau ^{+}-\tau ^{-}$ between the torque
recorded for increasing and decreasing field at $T=60$ mK with a
linear fit in the reversible region. The inset shows how $B_{{\rm
irr}}^{({\rm a})}$ can be found, helping the eye with a straight
line through the first points in the irreversible region.}
\label{fig:2}
\end{figure}

The measured torque loops shown in Fig. 1(a) clearly show that the
irreversible behavior vanishes quite quickly as the temperature
increases. The corresponding magnetization loops shown in Fig.
1(b) have been plotted only for $B_{{\rm a}}>0.2$ T, because the
division of the torque by the field results in uncertainties for
$B\approx 0$. Both torque and magnetization loops have a typical
shape that scales quite well with temperature. None of our data
shows jumps, peaks or fishtail effects. In principle $B_{{\rm
irr}}^{({\rm a})}$ could be easily determined as the point where
the branches for increasing and decreasing field first touch, but
we used a more accurate method to obtain $B_{{\rm irr}}^{({\rm
a})}$ as illustrated in Fig. 2. First, for the difference $\tau
^{+}-\tau ^{-}$ of the torque measurements between increasing and
decreasing field, respectively, we make a linear fit of what is
likely to be the reversible region. Then we define $B_{{\rm
irr}}^{({\rm a})}$ as the point where $\tau ^{+}-\tau ^{-}$
deviates from the fit, helping the eye with a straight line
through the irreversible data. For an ideal measurement, the
linear fit of $\tau ^{+}-\tau ^{-}$ in the reversible region
should be a constant equal to zero. However, because the
capacitance of the torquemeter is slightly temperature dependent,
the thermal drift leads to a non-zero slope for $\tau ^{+}-\tau
^{-}$. With the thermal stability of our experiments, that slope
is actually extremely small (see the scales of the inset in Fig.
2). But also the irreversible signal vanishes very smoothly for
increasing field, and neglecting the nonzero slope would lead to a
much higher uncertainty in the evaluation of $B_{{\rm irr}}^{({\rm
a})}$. Furthermore, this procedure becomes quite useful for the
measurements close to $T_{{\rm c}}$, where the signal-to-noise
ratio gets worse. Already above $T\approx 3$ K the irreversible
torque signal, although clearly present, becomes so small that a
reliable determination of $B_{{\rm irr}}^{({\rm a})}$ is no longer
possible. Above $T\approx 4$ K we found no more signs of
hysteretic magnetic behavior.

\begin{figure}[t]
\epsfxsize=8.5cm \centerline{\epsffile{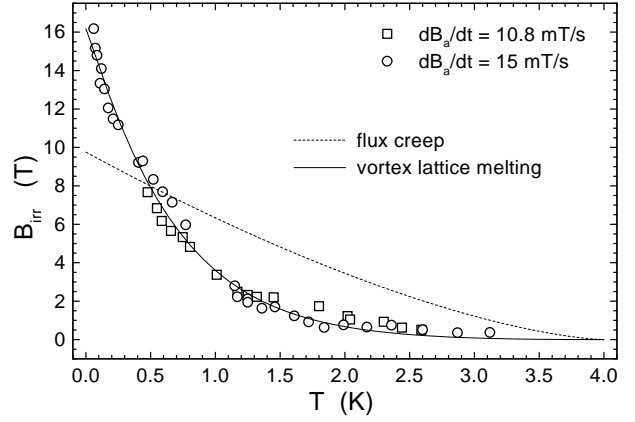}} \vspace{2mm}
\caption{The irreversibility line $B_{{\rm irr}}(T)$ of the
Bi-2201 single
crystal, recalculated for ${\bf B}\parallel c$ using $B_{{\rm irr}}=B_{{\rm %
irr}}^{({\rm a})}\cos 30^{\circ }$. The data points have been
obtained for different sweep rates as indicated. The dashed line
represents the flux-creep model (Eq. (1) with $n=1.5$), while the
solid line is a fit through the data for the vortex melting
transition using Eq. (2) with deduced $c_{{\rm L}}=0.13$.}
\label{fig:3}
\end{figure}

Fig. 3 shows the irreversibility line $B_{{\rm irr}}(T)$ of our
Bi-2201 sample, with a good overlap between the independent sets
of
measurements at d$B_{{\rm a}}/$d$t=15$ mT/s (resistive magnet) and at d$B_{%
{\rm a}}/$d$t=10.8$ mT/s (superconducting magnet). We tried to fit our data
with the very common function
\begin{equation}
B_{{\rm irr}}(T)=B_{{\rm irr}}(0)(1-T/T_{{\rm c}})^n  \eqnum{1}
\label{creep}
\end{equation}
keeping $T_{{\rm c}}=4$ K as a fixed parameter. This function has proved to
describe many experimental data with $n\sim 1.3\div 1.5$ and can be
qualitatively justified in a flux creep picture \cite{yesh88}, predicting $%
n=3/2$ or $4/3$ (the value of $n$ depends on the approximations
used to evaluate the pinning energy). In our case, a fit of the
data using Eq. (1)
with $n=1.5$ is not possible, while a good behavior is only attained for $%
n=5.2$ (not shown) which has no physical meaning. This is not
surprising, because Eq. (1) is obtained by supposing the flux
motion to take place for thermal activation over bulk pinning
barriers. Such a model works quite well in disordered systems like
Ba$_{1-x}$K$_x$BiO$_3$ \cite{goll96} but it is less suitable for
very clean superconductors. Actually, the vanishing of the
irreversible magnetization can also be ascribed to the melting of
the vortex solid \cite{farr96}. For a 3D anisotropic vortex
lattice, an useful form of the melting line $B_{{\rm m}}(T)$
\cite{blat94,houg89} has been obtained making use of the Lindemann
criterion, i.e. assuming that the lattice melts when the
mean-squared amplitude of the thermal vortex fluctuations $\langle
u^2\rangle _{th}$ exceeds a certain fraction $c_{{\rm L}}$ of the
vortex spacing. The melting line takes the form
\begin{equation}
B_{{\rm m}}(T)=B_{{\rm c2}}(0)\frac{4\vartheta ^2}{\left( 1+\sqrt{%
1+4\vartheta T_{{\rm s}}/T}\right) ^2}  \eqnum{2}  \label{melt}
\end{equation}
where $\vartheta =c_{{\rm L}}^2\sqrt{\beta _{{\rm m}}/Gi}(T_{{\rm c}}/T-1)$,
$T_{{\rm s}}=T_{{\rm c}}c_{{\rm L}}^2\sqrt{\beta _{{\rm m}}/Gi}$, $c_{{\rm L}%
}$ is the Lindemann number, $Gi=\frac 12\left( \frac{\gamma k_{{\rm B}}T_{%
{\rm c}}}{(4\pi /\mu _0)B_{{\rm c}}^2(0)\xi ^3(0)}\right) ^2$ is the
Ginzburg number, $\gamma =\sqrt{m_c/m_{ab}}$ is the anisotropy and $\beta _{%
{\rm m}}\approx 5.6$ is a numerical factor. This expression is
supposed to be valid over a wide temperature range below $T_{{\rm
c}}$, since it is calculated taking into account the suppression
of the order parameter close to $B_{{\rm c2}}$. The fit shown in
Fig. 3 is made fixing $T_{{\rm c}}=4$ K and leaving $B_{{\rm
c2}}(0)$ and $c_{{\rm L}}^2\sqrt{\beta _{{\rm m}}/Gi}$ as free
parameters. From our analysis we obtain $B_{{\rm c2}}(0)=16.4$ T
(yielding $\xi (0)=45$ \AA) and $c_{{\rm L}}^2\sqrt{\beta _{{\rm
m}}/Gi}=0.221$. Estimating $\kappa \sim 40$ and taking $\gamma
=350$ \cite{mart90,ando97} we find $Gi=3.3\cdot 10^{-2}$, and we
finally obtain $c_{{\rm L}}=0.13$.

It is worth noting that Hikami {\it et al.} \cite{hika91} have
studied the melting of a 3D flux lattice in strong magnetic
fields, obtaining a
criterion which is equivalent to the Lindemann's one with $c_{{\rm L}}=0.14$%
. This could explain why Eq. (2) gives a good description of the
data down to the lowest temperatures, i.e. up to the highest
fields. Eq. (2) takes into account only the contribution of
thermal fluctuations. We can neglect the contribution due to
quantum fluctuations of the vortices \cite{blat93}, because its
relative strength is given by $Q\propto \tilde Q/\sqrt{Gi}$, where
$\tilde Q=\frac{e^2\rho _{ab}}{\hbar d}$ ($d$ is the interlayer
spacing). Contrary to $Gi$, $\tilde Q$ is unaffected by the
anisotropy, so in Bi-2201 we expect the thermal fluctuations to be
much more enhanced than the quantum ones. Actually, the
contribution of quantum fluctuations of vortices should result in
a shift of the melting transition towards lower temperatures and
fields, which does not agree with the shape of our measured
$B_{{\rm irr}}(T)$.

With respect to the observed irreversible behavior, one should
notice that the torque and magnetization data shown in Fig. 1 tend
to rule out the existence of transitions between different phases
of the vortex solid, since at least for $T/T_{{\rm c}}<0.7$ no
jumps or fishtail effects are present (see e.g. \cite{khay96}).
From another point of view, it follows that no crossovers between
different pinning mechanisms are present. For the torque
magnetometry technique it can be shown \cite{qvar92} that the
pinning force density $F_{{\rm p}}(B)$ is proportional to the
hysteresis of the torque loop (i.e. what we called $\tau
^{+}-\tau^{-}$). In Fig. 4 we have plotted the field dependence of
the pinning force density at different temperatures, having
rescaled $F_{{\rm p}}$ by its maximum value $F_{{\rm p}}^{(\max
)}$ and the applied field by the measured $B_{{\rm irr}}^{({\rm
a})}$. All the curves tend to collapse into an unique shape, which
is also an indication that the magnetic field corresponding to the
pinning force maximum has approximately the same temperature
dependence as $B_{{\rm irr}}(T)$.

\begin{figure}[t]
\epsfxsize=8.5cm \centerline{\epsffile{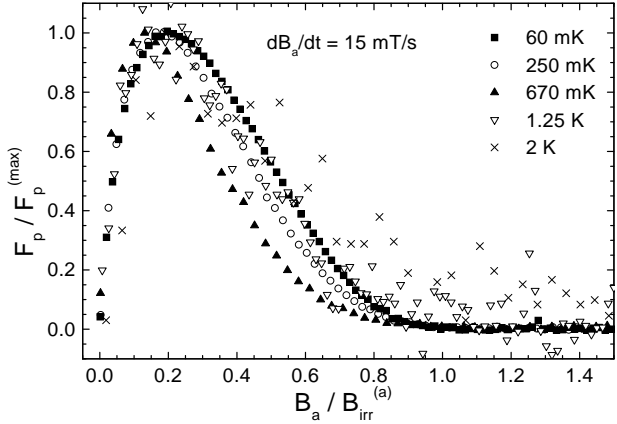}} \vspace{2mm}
\caption{Pinning force density $F_{{\rm p}}\propto \tau
^{+}-\tau^{-}$, normalized to the maximum value $F_{{\rm
p}}^{(\max )}$, at various temperatures as function of the
rescaled field $B_{{\rm a}}/B_{{\rm irr}}^{({\rm a})}$. The good
overlap and the very similar shapes of all the curves suggest the
absence of crossovers between different pinning mechanisms.}
\label{fig:4}
\end{figure}

From a qualitative point of view, the irreversibility line we measured can
be interestingly compared to the resistive critical field measured by
Osofsky {\it et al.} \cite{osof93} on a Bi-2201 thin film. The upward
curvature of the critical field obtained in this transport experiment is
very different from the saturating low-temperature behavior of $B_{{\rm c2}%
}(T)$ of the Werthamer-Helfand-Hohenberg theory \cite{wert66}. The
temperature dependence of the resistively determined critical
field is very similar to the irreversibility field reported here,
which confirms that flux lattice melting plays a crucial role in
the magnetoresistive transitions.

In conclusion, we have measured the irreversibility line of a
Bi-2201 single crystal down to $T=60$ mK and up to $B_{{\rm
a}}=28$ T, obtaining a curve that can be fitted with the form
predicted by the Lindemann criterion for the melting of the vortex
lattice. The magnetization loops do not show any jump or peak
effect, and the pinning force maintains the same shape as function
of the field throughout the investigated temperature range.
Finally, the behavior of $B_{{\rm irr}}(T)$ obtained here is very
similar to the resistive critical field of a Bi-2201 thin film,
suggesting that magnetoresistive experiments are likely to be
strongly influenced by flux lattice melting.

One of us (S.I.V.) was partially supported by the Russian Foundation for
Basic Research (Project no. 99-02-17877).

\end{document}